\documentclass[12pt]{iopart}

\usepackage{epsfig}

\begin{document}

\title{A two-variable series for the contact process with diffusion}

\author{W.G. Dantas and M.J. de Oliveira}

\address{Instituto de F\'{\i}sica,
Universidade de S\~{a}o Paulo,
Caixa Postal 66318
05315-970 S\~{a}o Paulo, S\~{a}o Paulo, Brazil}

\author{J.F. Stilck}
\address{Instituto de F\'{\i}sica, Universidade Federal Fluminense,
24210-340, Niter\'oi, RJ, Brazil}

\begin{abstract}
In this work we use the technique of the partial differential 
approximants to determine, from a pertubative 
supercritical series expansion for the ulimate survival probability, 
the critical line of the contact process model in one dimension with
diffusion and  
estimate the value of the crossover exponent that characterizes
the change of the critical behavior from the 1d directed percolation 
universality class to the mean-field directed percolation universality
class. This crossover occurs in the limit of infinite diffusion rate.  
\end{abstract} 


\maketitle

\section{Introduction}
In the last years there has been a growing interest in nonequilibrium
phase transitions \cite{md99,priv97}. 
The absence of a general theory
nonequilibrium systems originates to a number of 
open problems, even in one-dimensional systems 
that, in the equilibrium regime, generally are exactly solvable.
Usually, numerical simulations are a userful
technique in the study of phase transitions and critical
phenomena, and its power has been growing with the increasing capacity
of the computers and the development of new simulational
techniques. Non-equilibrium models are particularly suited for
simulations. However, other approximate approaches may be complementary
in the study of these phenomena.
Thus, it is of interest to study the systems by other techniques. 
One of the successful techniques 
are the series expansions \cite{dj91}, which in some cases
lead to very precise estimates 
for the critical properties that characterize 
these transitions \cite{dj91,dickjenss}.

Among the non-equilibrium models, are the ones that exhibit {\em
  absorbing} states, that is, states that may be reached in their 
dynamics, but transitions leaving them are forbidden. Since such
models do not obey detailed balance, they are intrinsically out of
equilibrium. 
The most studied system  for this class of problems is the so called 
contact process (CP) model \cite{h74}, a {\em toy model} for the 
spreading of an epidemic. This model displays a transition 
between an absorbing and an active state with critical exponents
belonging to the directed percolation (DP)
universality class \cite{dpjensen}. In addition, the CP 
model is related to the Schl\"ogl's lattice model for autocatalytic 
reactions \cite{s72} and the Reggeon Field Theory \cite{gt}.      

Many variants of this model have been studied \cite{dicktom91,mariofiore,
wgdstilckcross,yoon05}, most of them 
belonging to DP class also.  In fact, the 
robustness of this universality class is an evident characteristic of these
models. Such robustness is explained by the conjecture that 
all models with phase transitions between active and absorbing states 
with a scalar order parameter, short range interactions and no 
conservation laws belong to this class \cite{j81}.

One of these variants is the CP with diffusion \cite{dickjensdif},
which exhibits  
a critical line. This line begins at the critical point of the model 
without diffusion and ends in 
the infinite diffusion rate limit, where the critical properties of
the system  
approaches those predicted by the mean-field approximation. The
mean-field behavior of the model in the limit of infinite diffusion
rate may be understood considering that, since diffusion processes are
dominating the evolution of the system in this limit, creation
processed are effectively determined by the mean densities, as is
supposed in the mean-field approximation. 
This change of behavior at the infinite diffusion rate limit, 
between the critical behavior of the DP class and the one predicted by 
the mean-field approximation, characterizes a crossover of the
critical properties in  
the neighborhood of a multicritical point.  As in the equilibrium case
we may then write any density variable, in the neighborhood of a
multicritical point, as the following scaling form \cite{fk77}:
\begin{eqnarray}
g(\lambda,D)=(\lambda_c-\lambda)^{\theta}F\left(
\frac{D_c-D}{|\lambda_c-\lambda|^{\phi}}\right),
\end{eqnarray}
where $\lambda$ is a transition rate of the original CP model,
$D$ is the diffusion rate, ranging between 0 and 1, $\theta$ is 
a critical exponent associated
to the density variable $g$, 
corresponding to the value predicted by the mean-field approximation.
The scaling function 
$F(z)$ is singular at a point $z=z_0$ of its argument, such that the
critical line, 
in the neighborhood of the multicritical point, corresponds to
\begin{eqnarray}
(D_c-D)=z_0(\lambda_c-\lambda)^{\phi},
\end{eqnarray}
where $\phi$ is the crossover exponent, and the critical exponent of
$g$ on this line is determined by this singularity, being in general
different from $\theta$.

One of the first studies of this problem was performed by
Dickman and Jensen \cite{dickjensdif}, 
who considered the model using   
supercritical series in $\lambda$ with the diffusion rate $D$ taken 
as a fixed parameter. Therefore, in their calculation series
expansions are derived for fixed values of the diffusion rate $D$, and
analysing the series leads, among other information, to the phase
diagram of the model with diffusion. 
However, they found that the fluctuations of the estimates provided by 
d-log Pad\'e approximants show increasing fluctuations as the
diffusion rate grows, so that the critical curve was obtained only up to 
$D\approx 0.8$. Since the crossover exponent $\phi$ characterizes the
critical curve close to the infinite diffusion rate limit $D \to
\infty$ no precise estimate of the crossover exponent was
possible. The disapointing performance of the Pad\'e approximants as
the multicritical point is approached is not surprising, since it is
known tha one-variable series analysis techniques are not effective
close to such points \cite{fk77}.
More recently \cite{mariofiore}, the model was simulated made 
in a the ensemble where the number of particles is conserved. These
simulations display smaller  
fluctuations in the estimates, enabling the estimation of the critical 
line up to values near to the multicritiacal point, an leading 
the value $\phi=4.03(3)$ for the crossover exponent.

This result is consistent with the lower bound $\phi\geq 1$
predicted by Katori in \cite{kato94}.

In the present work, we obtain estimates for the critical line and the
crossover exponent $\phi$ 
using a two-variable supercritical series analyzed 
using partial differential approximants (PDA´s) 
\cite{fk77,s90,stilcksalinas}. This technique seems to be more
appropriate for the analysis of a  
two-variable series with a multicritical behavior, as shown by the
results obtained from other models \cite{wgdstilckcross,adlersalman}.     
This paper is organized as follows. In section II we present 
the model and the mean-field results, in section III we show
the derivation of the supercritical series and in section IV 
the analysis of this series is presented. Finally, in section V we
the conclusions and final discussions of this work may be found.  

\section{Definition of the model and mean-field results}

\begin{figure}
\begin{center}
\vspace*{1.1cm}
\epsfig{file=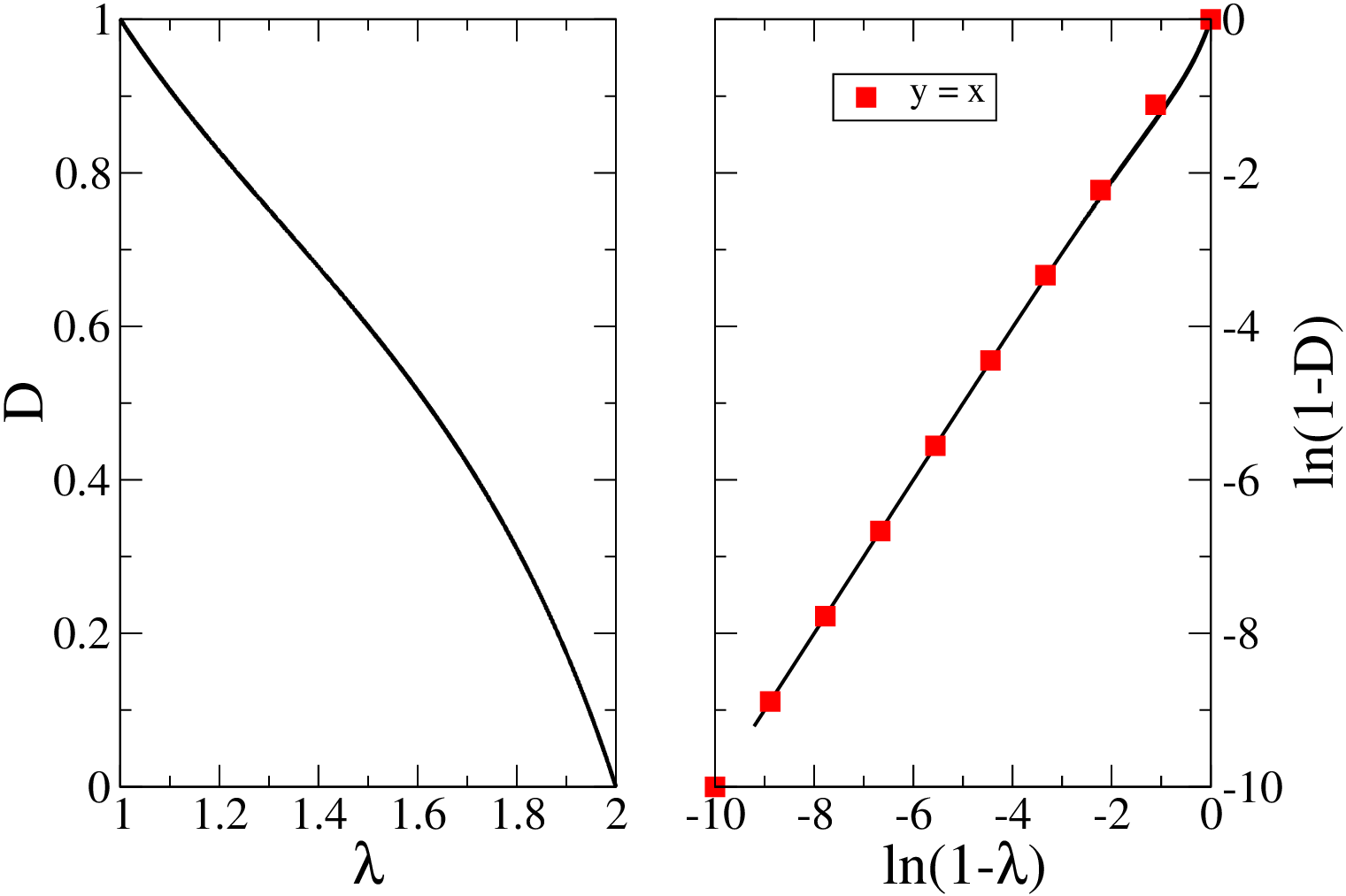,scale=0.5}
\vspace{0.1 cm}
\caption{Left: phase diagram obtained
using the two-site mean-field approach. Right: the log-log plot of the
same quantity is plotted and compared to the result $\phi=1$ for this
approach.} 
\label{fig1}
\end{center}
\end{figure}

In a $d$-dimensional lattice each site can be empty or 
occupied by a particle, so that we will associate an occupation
variable $\eta_i=0,1$ to the site $i$.  
The evolution of the system is governed by
markovian local rules such that the particles are annihilated
with rate $1$ and created in 
empty sites with a transition rate
$\lambda n/z$, where $\lambda$ is a positive parameter, 
$n$ is the number of occupied nearest neighbors and $z$ 
is the total number of nearest neighbors.
In addition to these rules that define the CP, 
we include a diffusve process, allowing 
the hopping of particles to empty nearest neighbour sites 
at the rate $\tilde{D}=D/(1-D)$. The configuration such that all sites 
are empty is an absorbing state. The transition from an active
steady state, with a nonzero density of particles, to
the absorbing state defines a transition line in the $(\lambda,D)$
plane as shown in figure \ref{fig1}.

A mean-field approach for this model can be obtained at several
levels of approximation \cite{md99}.  In the one-site level
the role of the diffusion is irrelevant since 
it contributed equally to creation and annihilation of
particles at a given site $i$.  Already in the 
two-site level it is possible to determine the critical line 
by using as variables the parameters 
$\lambda$ and $\tilde{D}$. This line is described 
by the expression
\begin{eqnarray}
\lambda_c=\frac{1}{2}\left[2-\tilde{D}+\sqrt{\tilde{D}^2+4}\right]
\end{eqnarray}
and it is not difficult to show that in the neighborhood of 
the multicritical point, $(\lambda_c=1,\tilde{D}=1)$, the behavior of
this curve is given by the scaling relation:
\begin{eqnarray}
(1-D)\sim(\lambda-1)^{\phi}
\end{eqnarray}
where $\phi=1$. This crossover behavior may be seen in figure \ref{fig1}.

This result is in accordance with the lower bound, $\phi\geq 1$,
determined by Katori \cite{kato94}.
But, as is well known , the mean-field
approach always overestimates the supercritical
region of the models so that it is 
not surprising that more precise results for the
exponent $\phi$ will differ from this unitary value.
In the next section we will derive the 
supercritical series in the variables $\lambda$ and $\tilde{D}$
to determine the value of this exponent and compare it
with that obtained in \cite{mariofiore,dickjensdif}. 

\section{Derivation of the supercritical series for the model}
\label{series}

We use the operator formalism proposed 
by Dickman and Jensen \cite{dickjenss} in order to
derive a supercritical series. To this end, we define the 
microscopic configuration of the lattice, $|\eta\rangle$, 
as the direct product of kets
$| \eta \rangle= \bigotimes_i | \eta_i \rangle$,
with the following orthonormality property,
$\langle \eta | \eta^\prime \rangle = \prod_i \delta_{\eta_i,\eta_i\prime}$.
The particle creation and annihilation operators at site
$i$ are defined as 
\begin{eqnarray}
A_i^\dagger |\eta_i \rangle &=&(1-\eta_i)|\eta_i+1\rangle,\nonumber \\
A_i |\eta_i\rangle &=& \eta_i|\eta_i-1\rangle.
\end{eqnarray}
In this formalism, the state of the system at time $t$ is
\begin{equation}
|\psi(t) \rangle = \sum_{\{\eta\}} p(\eta,t) |\eta \rangle,
\end{equation}
where $p(\eta,t)$ is the probability of a configuration $\eta$ at time $t$.
If we define the projection onto all possible states as
$\langle \; | \equiv \sum_{\{\eta\}} \langle\eta|$
then the normalization of the state of the system may be expressed as
$\langle \; |\psi \rangle =1$. In this notation, the master equation
for the evolution of the state is:
\begin{equation}
\frac{d |\psi(t)\rangle}{d t}=S|\psi(t)\rangle.
\label{me1}
\end{equation}
The evolution operator $S$ may be expressed in terms of the creation and
annihilation operators as $S=\mu R + V$ where
\begin{eqnarray}
\label{eqops}
R&=&\tilde{D}\sum_i (1-A_{i-1}^{\dagger}A_i)A_{i-1}A_i^{\dagger}+(1-A_{i+1}^{\dagger}A_i)
A_{i+1}A_i^{\dagger}+\nonumber\\
&+&\sum_i(A_i-A_i^{\dagger}A_i), \\
\label{eqopsv}
V &=& \sum_i (A_i^\dagger - A_iA_i^\dagger)(A_{i-1}^\dagger A_{i-1} +
A_{i+1}^\dagger A_{i+1}),
\end{eqnarray}
where $\mu\equiv 2/\lambda$.

We notice that the operator $R$ diffuses $(01\to 10)$ or annihilates 
particles $(1\to 0)$, while the operator 
$V$ acts in the opposite way, generating particles $(0 \to 1)$. 
It is convenient to join the diffusion with the annihilation process 
to avoid ambiguities in the truncation of the series at a certain order.
For small values of the parameter $\mu$ the
creation of particles is favored, and the decomposition above is convenient
for a supercritical perturbation expansion. Using the equations 
(\ref{eqops}) and (\ref{eqopsv}) the action of each operator on a generical 
configuration $(\mathcal{C})$ is given by 
\begin{eqnarray}
R({\mathcal{C}})&=&\tilde{D}\left[\sum_i({\mathcal{C}}^{\prime}_i)+
\sum_j({\mathcal{C}}^{\prime r}_j)+
\sum_k({\mathcal{C}}^{\prime l}_k)\right]+\nonumber\\
&+&\sum_t({\mathcal{C}}^{\prime\prime}_t)+
-[(r_1+2r_2)+\tilde{D}+r]({\mathcal{C}}),
\label{as0}
\end{eqnarray}
where the first sum is over $r_1$ sites with particles 
and one empty neighbor, the two next sums are over $r_2$ sites with 
particles and two empty neighbors and the last sum is over
all sites occupied by a particle.  Configuration $({\mathcal{C}}^{\prime}_i)$ 
is obtained moving the particle at the site $i$ to the single empty
neighbor site,
$({\mathcal{C}}^{\prime (r,l)}_i)$ is a configuration where the particle
at the site $i$ moved to the empty neighbor at the right ($r$) or at
the left ($l$)is replaced by a hole and one of the empty neighbors (at the
right or left).
In the other hand, the action of operator $V$ is
\begin{equation}
V({\mathcal{C}})=\sum_i ({\mathcal{C}}^{\prime \prime\prime}_i)+2 \sum_j
({\mathcal{C}}^{\prime \prime\prime}_j) -(q_1+2q_2)({\mathcal{C}}),
\label{av}
\end{equation}
where the first sum is over the $q_1$ empty sites with one occupied
neighbor, the second sum is over the $q_2$ empty sites with two occupied
neighbors. Configuration $({\mathcal{C}}_i^{\prime \prime\prime})$ is obtained
occupying the site $i$ in configuration $(\mathcal{C})$.

To obtain a supercritical expansion for the ultimate survival probability of 
particles, we start by remembering that in order to access the long-time
behavior of a quantity, it is useful to consider its Laplace transform,

\begin{equation}
|\tilde{\psi}(s) \rangle = \int_0^\infty e^{-st} |\psi(t)\rangle.
\label{lt}
\end{equation}
Inserting the formal solution $|\psi(t)\rangle =e^{St} |\psi(0)\rangle$ in
the master equation (\ref{me1}) we find
\begin{equation}
|\tilde{\psi}(s) \rangle = (s-S)^{-1} |\psi(0)\rangle.
\label{tpsi}
\end{equation}
The stationary state $|\psi(\infty) \rangle \equiv \lim_{t \to \infty} |\psi(t)
\rangle$ may then be found by means of the relation 
\begin{equation}
|\psi(\infty) \rangle = \lim_{s \to 0} s |\tilde{\psi}(s) \rangle.
\end{equation}
A perturbative expansion may be obtained by assuming that
$|\tilde{\psi}(s) \rangle$  
can be expressed in powers of $\mu$ and using (\ref{tpsi}),
\begin{equation}
|\tilde{\psi}(s) \rangle = |\tilde{\psi}_0 \rangle+\mu |\tilde{\psi}_1 
\rangle +\mu^2 |\tilde{\psi}_2 \rangle + \cdots = (s- V -\mu R)^{-1}
|\psi(0) \rangle.  
\end{equation}
Since
\begin{equation}
(s- V -\mu R)^{-1}= (s-V)^{-1} \left[ 1 + \mu (s-V)^{-1}R 
+ \mu^2 (s-V)^{-2} R^2 + \cdots \right],
\end{equation}
we arrive at
\begin{equation}
|\tilde{\psi}_0 \rangle = (s-V)^{-1} |\psi(0)\rangle
\end{equation}
and
\begin{equation}
|\tilde{\psi}_n \rangle = (s-V)^{-1} R |\tilde{\psi}_{n-1} \rangle,
\end{equation}
for $n\geq 1$.
The action of the operator $(s-V)^{-1}$ on an arbitrary configuration
$({\mathcal{C}})$ may be found by noticing that
\begin{equation}
(s-V)^{-1} ({\mathcal{C}})=s^{-1}\left\{({\mathcal{C}})+ (s-V)^{-1}V
({\mathcal{C}})\right\}, 
\end{equation}
and using the expression \ref{av} for the action of the operator $V$, we get
\begin{equation}
(s-V)^{-1} ({\mathcal{C}})= s_q \left\{({\mathcal{C}}) + (s-V)^{-1} \left[
\sum_i 
({\mathcal{C}}^{\prime\prime\prime}_i)+2 \sum_j
({\mathcal{C}}^{\prime\prime\prime}_j) 
\right] \right\},
\label{sv}
\end{equation}
where the first sum is over the $q_1$ empty sites and one occupied
neighbor, the second sum is over the $q_2$ empty sites and two occupied
neighbors, and we define $s_q \equiv 1/(s+q_1+2 q_2)$.

It is convenient to adopt as the initial configuration a translational
invariant one with a single particle (periodic boundary conditions are
used). Now, looking at the recursive expression (\ref{sv}), we may notice 
that the operator $(s-V)^{-1}$ acting on any configuration generates 
an infinite set of configurations, and thus we are unable to calculate 
$|\tilde{\psi}\rangle$ in a closed form. However, it is possible calculate the 
extinction probability $\tilde{p}(s)$, which corresponds to the coefficient of 
the vacuum state $|0\rangle$. As happens also for  models
\cite{dickjenss,dj91}   
related to the CP, configurations with more than $j$ particles 
only contribute at orders higher than $j$, and since we are interested
in the ultimate survival 
probability for particles $P_\infty=1-\lim_{s \to 0} s \tilde{p}(s)$, $s_q$
may be replaced by $1/q$ in equation (\ref{sv}). An illustration of
this procedure may be found in a previous calculation \cite{wgdstilckcross}.

The algebraic operations described above is performed by a simple algorithm
wich permit us to calculate 24 terms with a processing time to about
2 hours. Actually, the limiting factor in this operation is the memory 
required. In this way we define the coefficients $b_{i,j}$ for the ultimate
survival probability as:

\begin{eqnarray}
\label{pinf}
P_{\infty}=1-\frac{1}{2}\mu-\frac{1}{4}\mu^2-\sum_{i=3}^{24}\sum_{j=0}^{i-2}
b_{i,j}\mu^i\tilde{D}^j,
\end{eqnarray}
and they are listed in Table \ref{tab2}.

\begin{table}
\caption{Coefficients for the series expansion for ultimate survival
probability corresponding to the CP model with diffusion. The indexes
refer them to the equation (\ref{pinf}).}

\smallskip
\begin{tabular}{cccccc} 
\hline
\emph{i} & \emph{j} &$b_{i,j}$&\emph{i} & \emph{j} &$b_{i,j}$  \\
\hline
3 &0& 0.25000000000000000000$\times 10^0$      & &6&  0.10567643059624561630$\times 10^2$\\
  &1&-0.25000000000000000000$\times 10^0$      & &7&  0.34998474121093847700$\times 10^1$\\
4 &0& 0.28125000000000000000$\times 10^0$      &  &8& 0.12568359375000000000$\times 10^2$\\
  &1&-0.37500000000000000000$\times 10^0$      &11&0& 0.24775957118666096513$\times 10^1$\\
  &2& 0.37500000000000000000$\times 10^0$      &  &1&-0.71703082845177412707$\times 10^1$\\
5 &0& 0.34375000000000000000$\times 10^0$      &  &2& 0.12429764028158224676$\times 10^2$\\
  &1&-0.50781250000000000000$\times 10^0$      &  &3&-0.16676337106809967281$\times 10^2$\\   
  &2& 0.57812500000000000000$\times 10^0$      &  &4& 0.19435605471721252968$\times 10^2$\\
  &3&-0.62500000000000000000$\times 10^0$      &  &5&-0.15230894658300556443$\times 10^2$\\  
6 &0& 0.44726562500000000000$\times 10^0$      &  &6& 0.13923689787279895924$\times 10^2$\\
  &1&-0.76220703125000000000$\times 10^0$      &  &7&-0.24910518081099844778$\times 10^2$\\
  &2& 0.85058593750000000000$\times 10^0$      &  &8&-0.13853664539478481643$\times 10^2$\\
  &3&-0.83984375000000000000$\times 10^0$      &  &9&-0.23740234375000000000$\times 10^2$\\
  &4& 0.10937500000000000000$\times 10^0$\\
7 &0& 0.60223388671874955591$\times 10^0$      &12&0& 0.36488812342264926869$\times 10^1$\\
  &1&-0.11734619140625004441$\times 10^1$      &  &1&-0.11443929729648042226$\times 10^2$\\
  &2& 0.15190429687499997780$\times 10^1$      &  &2& 0.21418735868689896762$\times 10^2$\\  
  &3&-0.14140624999999984457$\times 10^1$      &  &3&-0.29831307350977681381$\times 10^2$\\
  &4& 0.10878906249999982236$\times 10^1$      &  &4& 0.34272964785342651339$\times 10^2$\\ 
  &5&-0.19687500000000000000$\times 10^1$      &  &5&-0.40398672142671166796$\times 10^2$\\
8 &0& 0.83485031127929687500$\times 10^0$      &  &6& 0.27855684964608855125$\times 10^2$\\
  &1&-0.18110389709472716202$\times 10^1$      &  &7&-0.13595902518316915319$\times 10^2$\\ 
  &2& 0.25234603881835981909$\times 10^1$      &  &8& 0.60935236387946176251$\times 10^2$\\ 
  &3&-0.29291381835937464473$\times 10^1$      &  &9& 0.40270442479922508028$\times 10^2$\\
  &4& 0.24864501953125062172$\times 10^1$      &  &10&0.45106445312500000000$\times 10^2$\\
  &5&-0.10754394531250017764$\times 10^1$      &13&0& 0.54293656084851154020$\times 10^1$ \\ 
  &6& 0.36093750000000000000$\times 10^1$      &  &1&-0.18322144692814863021$\times 10^2$\\
9 &0& 0.11814667913648828623$\times 10^1$      &  &2& 0.36259195896082665911$\times 10^2$\\
  &1&-0.28569926950666579835$\times 10^1$      &  &3&-0.54866931326313050477$\times 10^2$\\
  &2& 0.42781094621729156557$\times 10^1$      &  &4& 0.67130818799164941879$\times 10^2$\\
  &3&-0.48761836864330092567$\times 10^1$      &  &5&-0.64293858561553619779$\times 10^2$\\
  &4& 0.54410674483687788694$\times 10^1$      &  &6& 0.81638245065997452343$\times 10^2$\\
  &5&-0.48496839735243062464$\times 10^1$      &  &7&-0.59206203158103356543$\times 10^2$\\
  &6& 0.11848958333333414750$\times 10^0$      &  &8&-0.11386872839957611347$\times 10^2$\\
  &7&-0.67031250000000000000$\times 10^1$      &  &9&-0.15017302299755911577$\times 10^3$\\
10&0& 0.16988672076919952847$\times 10^1$      &  &10&-0.10358591484729181786$\times 10^3$\\
  &1&-0.45030223008843064392$\times 10^1$      &  &11&-0.8611230468750000000$\times 10^2$\\
  &2& 0.73700355965291270977$\times 10^1$      &14&0&0.8132542219307161701600$\times 10^1$\\
  &3&-0.92486491500105252328$\times 10^1$      &  &1&-0.29467694610727896531$\times 10^2$\\
  &4& 0.87502182305104447835$\times 10^1$      &  &2& 0.62075441392908530247$\times 10^2$\\
  &5&-0.93882905701060082038$\times 10^1$      &  &3&-0.96520364146752442025$\times 10^2$\\
\end{tabular}
\label{tab2}
\end{table}

\begin{table}
\smallskip
\begin{tabular}{cccccc} 
\hline
\emph{i} & \emph{j} &$b_{i,j}$&\emph{i} & \emph{j} &$b_{i,j}$  \\
\hline
14&4& 0.12740413805065173847$\times 10^3$       &  &3&-0.55737662171810006839$\times 10^3$\\
  &5&-0.14872799806680012580$\times 10^3$       &  &4& 0.80584299514354984240$\times 10^3$\\ 
  &6& 0.11055259565281477308$\times 10^3$       &  &5&-0.10663609887965685630$\times 10^4$\\ 
  &7&-0.15322889380857196784$\times 10^3$       &  &6& 0.12359653482806177180$\times 10^4$\\
  &8& 0.15125687744198603468$\times 10^3$       &  &7&-0.86967640846259655518$\times 10^3$\\
  &9& 0.12100775973170790678$\times 10^3$       &  &8& 0.16397080915531578285$\times 10^4$\\
  &10&0.36717258144235222517$\times 10^3$       &  &9&-0.72176857313912660175$\times 10^3$\\
  &11&0.24952145042880511028$\times 10^3$       &  &10&-0.39467789553376610456$\times 10^3$\\  
  &12&0.16504858398437500000$\times 10^3$       &  &11&-0.37241724510229601037$\times 10^4$\\
15&0& 0.12275012836144505002$\times 10^2$       &  &12&-0.43433579993156563432$\times 10^4$\\
  &1&-0.47363165128788978109$\times 10^2$       &  &13&-0.49111545112523144780$\times 10^4$\\
  &2& 0.10546586796137503939$\times 10^3$       &  &14&-0.28643360597728060384$\times 10^4$\\
  &3&-0.1762739398241103288$\times 10^3$        &  &15&-0.11834527587890625000$\times 10^4$\\
  &4& 0.23298609118631188153$\times 10^3$       &18&0& 0.435207828742268674200$\times 10^2$\\
  &5&-0.27071715385838172097$\times 10^3$       &  &1& -0.20009555228747112210$\times 10^3$\\
  &6& 0.33267266081610591755$\times 10^3$       &  &2& 0.51666085550451919062$\times 10^3$\\
  &7&-0.18014432368848466126$\times 10^3$       &  &3&-0.98221829678260564833$\times 10^3$\\
  &8& 0.24323790718884771422$\times 10^3$       &  &4& 0.15468813789798582548$\times 10^4$\\
  &9&-0.43351480900655008099$\times 10^3$       &  &5&-0.19452358996876919264$\times 10^4$\\
  &10&-0.48110863117407200207$\times 10^3$      &  &6& 0.23142581042234874076$\times 10^4$\\
  &11&-0.88513499744041246231$\times 10^3$      &  &7&-0.29269143532222028625$\times 10^4$\\
  &12&-0.57707323452079549497$\times 10^3$      &  &8& 0.11929249112512670763$\times 10^4$ \\
  &13&-0.31740112304687500000$\times 10^3$      &  &9&-0.33469260525748682085$\times 10^4$\\
16&0& 0.18620961415130427241$\times 10^2$       &  &10&0.22662397929431922421$\times 10^4$ \\
  &1&-0.76547748518027589171$\times 10^2$       &  &11&0.33534053058169365613$\times 10^4$\\
  &2& 0.17936794520034777634$\times 10^3$       &  &12&0.10497352179518215053$\times 10^5$\\
  &3&-0.30967915791812674797$\times 10^3$       &  &13&0.11603782689493993530$\times 10^5$\\
  &4& 0.45127497217248452444$\times 10^3$       &  &14&0.11326231062527707763$\times 10^5$ \\
  &5&-0.53698491310493750461$\times 10^3$       &  &15&0.62269615466431168898$\times 10^4$ \\
  &6& 0.51912309945306844838$\times 10^3$       &  &16&0.22929397201538085938$\times 10^4$ \\
  &7&-0.74776029427868388666$\times 10^3$       &19&0& 0.66930218067969633466$\times 10^2$ \\
  &8& 0.31390176074091152714$\times 10^3$       &  &1&-0.32354897975245813768$\times 10^3$\\
  &9&-0.23066668763466492464$\times 10^3$       &  &2& 0.88042629554806353553$\times 10^3$\\
  &10&0.12806582574012411442$\times 10^4$       &  &3&-0.17413898514485581472$\times 10^4$\\
  &11&0.15251186818533849419$\times 10^4$       &  &4& 0.27574673463423659996$\times 10^4$\\
  &12&0.21006816788719602300$\times 10^4$       &  &5&-0.39760107863198809355$\times 10^4$\\
  &13&0.12983815231244370807$\times 10^4$       &  &6& 0.45478180647223589403$\times 10^4$\\
  &14&0.61213073730468750000$\times 10^3$       &  &7&-0.44829538123020120111$\times 10^4$\\
17&0& 0.28405733950686048672$\times 10^2$       &  &8& 0.71439950475520599866$\times 10^4$\\
  &1&-0.12342415559750365617$\times 10^3$       &  &9&-0.11433220550468702186$\times 10^4$\\
  &2& 0.30541526863109334045$\times 10^3$       &  &10&0.58699584417021997069$\times 10^4$\\
\end{tabular}
\end{table}

\begin{table}
\smallskip
\begin{tabular}{cccccc} 
\hline
\emph{i} & \emph{j} &$b_{i,j}$&\emph{i} & \emph{j} &$b_{i,j}$  \\
\hline
19&11&-0.80007309950477119855$\times 10^4$        &  &15&-0.19550932821134978440$\times 10^6$\\
  &12&-0.13849586891129882133$\times 10^5$        &  &16&-0.17865808405461237999$\times 10^6$\\ 
  &13&-0.28630836746692133602$\times 10^5$        &  &17&-0.13001245697926016874$\times 10^6$\\ 
  &14&-0.29699369329023520550$\times 10^5$        &  &18&-0.60334798749708410469$\times 10^5$\\
  &15&-0.25808082841629722679$\times 10^5$        &  &19&-0.16853935146331787109$\times 10^5$\\
  &16&-0.13385541065918856475$\times 10^5$        &22&0&  0.24876519640902955643$\times 10^3$\\
  &17&-0.44510006332397460938$\times 10^4$        &  &1& -0.13849806980532957823$\times 10^4$\\
20&0&  0.10337399908883011790$\times 10^3$        &  &2&  0.42491525612070690840$\times 10^4$\\  
  &1& -0.52548922262251915072$\times 10^3$        &  &3& -0.95509490156905540061$\times 10^4$\\
  &2&  0.14837268484319015442$\times 10^4$        &  &4&  0.17072122607398174296$\times 10^5$\\
  &3& -0.30794482965317188246$\times 10^4$        &  &5& -0.25259991312968326383$\times 10^5$\\
  &4&  0.51825134723219762236$\times 10^4$        &  &6&  0.36959825267281092238$\times 10^5$\\
  &5& -0.70384028435684167562$\times 10^4$        &  &7& -0.39724946702482979163$\times 10^5$\\
  &6&  0.95392163017937873519$\times 10^4$        &  &8&  0.41082488201681495411$\times 10^5$\\
  &7& -0.10775090683332531626$\times 10^5$        &  &9& -0.69996295461125846487$\times 10^5$\\
  &8&  0.72306238600702890835$\times 10^4$        &  &10&-0.47433844045513687888$\times 10^4$ \\
  &9& -0.17556031650766155508$\times 10^5$        &  &11&-0.94472082453477501986$\times 10^5$\\
  &10& 0.45576565336850308086$\times 10^3$        &  &12& 0.18231631734823597071$\times 10^5$\\
  &11&-0.68853229355527937514$\times 10^4$        &  &13& 0.74507403911089320900$\times 10^5$ \\
  &12& 0.27725591978402942914$\times 10^5$        &  &14& 0.28127919612702319864$\times 10^6$\\
  &13& 0.46620453043310422800$\times 10^5$        &  &15& 0.40344708716105029453$\times 10^6$\\
  &14& 0.75787894589888033806$\times 10^5$        &  &16& 0.49346542769156675786$\times 10^6$ \\
  &15& 0.73699439769879478263$\times 10^5$        &  &17& 0.42523410507562680868$\times 10^6$\\
  &16& 0.58191226154708187096$\times 10^5$        &  &18& 0.28817317888963740552$\times 10^6$\\
  &17& 0.28519783989193914749$\times 10^5$        &  &19& 0.12690253004534545471$\times 10^6$ \\
  &18& 0.86547234535217285156$\times 10^4$        &  &20& 0.32865173535346984863$\times 10^5$\\
21&0&  0.15998830271612598608$\times 10^3$        &23&0&  0.38696067609131841891$\times 10^3$\\
  &1& -0.85206464313625656359$\times 10^3$        &  &1& -0.22521288931067451813$\times 10^4$ \\
  &2&  0.25270411365871177622$\times 10^4$        &  &2&  0.72239403631839231821$\times 10^4$  \\
  &3& -0.53883333002320678133$\times 10^4$        &  &3& -0.16540644823073282168$\times 10^5$  \\
  &4&  0.93069165541648508224$\times 10^4$        &  &4&  0.30965978110985601234$\times 10^5$ \\
  &5& -0.14256926805592749588$\times 10^5$        &  &5& -0.49715203648917136888$\times 10^5$\\
  &6&  0.16894039347875652311$\times 10^5$        &  &6&  0.63140021723358484451$\times 10^5$  \\
  &7& -0.21057424914324066776$\times 10^5$        &  &7& -0.90160716109902248718$\times 10^5$ \\
  &8&  0.2676311383571557235$\times 10^5$         &  &8&  0.96095356087432839558$\times 10^5$\\
  &9& -0.76173492536429366737$\times 10^4$        &  &9& -0.63690648044614848914$\times 10^5$\\
  &10& 0.42006937485571783327$\times 10^5$        &  &10& 0.18830794813510467065$\times 10^6$\\
  &11&-0.12698208721358355433$\times 10^4$        &  &11& 0.57163083968398816069$\times 10^5$\\
  &12&-0.53474107987358220271$\times 10^4$        &  &12& 0.19078924010583921336$\times 10^6$\\
  &13&-0.90835383726464555366$\times 10^5$        &  &13&-0.11462564856718564988$\times 10^6$\\
  &14&-0.14158471202173284837$\times 10^6$        &  &14&-0.34295621170633088332$\times 10^6$\\
\end{tabular}
\end{table}

\begin{table}
\smallskip
\begin{tabular}{cccccc} 
\hline
\emph{i} & \emph{j} &$b_{i,j}$&\mbox{ }&\mbox{ }&\mbox{ }\\
\hline
23&15&-0.83015840201854216866$\times 10^6$\\
  &16&-0.11002212525564364623$\times 10^7$\\
  &17&-0.12227985275158903096$\times 10^7$\\
  &18&-0.99733247054306510836$\times 10^6$\\
  &19&-0.63431234698974736966$\times 10^6$\\
  &20&-0.26563714369463571347$\times 10^6$\\
  &21&-0.64165338807344436646$\times 10^5$\\
24&0&  0.60509199550873199769$\times 10^3$\\ 
  &1& -0.36606232348859748527$\times 10^4$\\
  &2&  0.12139375556095224965$\times 10^5$\\
  &3& -0.29330923818458009919$\times 10^5$\\ 
  &4&  0.55510496259245075635$\times 10^5$\\
  &5& -0.89523794399456470273$\times 10^5$\\  
  &6&  0.13690056986421268084$\times 10^6$\\
  &7& -0.14950913041347730905$\times 10^6$\\
  &8&  0.20276027262469305424$\times 10^6$\\  
  &9& -0.24868958310934680048$\times 10^6$\\ 
  &10& 0.44658906685524416389$\times 10^5$\\ 
  &11&-0.50473239786133670714$\times 10^6$\\
  &12&-0.20583143731838543317$\times 10^6$\\
  &13&-0.31319042987920023734$\times 10^6$\\ 
  &14& 0.51903724071582528995$\times 10^6$\\
  &15& 0.12406518345377091318$\times 10^7$\\
  &16& 0.23558257104359627701$\times 10^7$\\  
  &17& 0.29049027840298512019$\times 10^7$\\  
  &18& 0.29832956761808455922$\times 10^7$\\
  &19& 0.23110049545479607768$\times 10^7$\\
  &20& 0.13877345160856433213$\times 10^7$\\
  &21& 0.55381177327087428421$\times 10^6$\\
  &22& 0.12541407130479812622$\times 10^6$\\
\hline
\end{tabular}
\end{table}

\section{Analysis of the series}
\label{analysis}

To obtain estimates of the critical properties, specially
the critical line, from the supercritical series 
for the ultimate survival probability as given by the 
equation (\ref{pinf}), we  initially use
d-log Pad\'e approximants. These approximants
are defined as ratios of two polynomials 

\begin{eqnarray}
F_{LM}(\lambda)=\frac{P_L(\lambda)}{Q_M(\lambda)}=
\frac{\sum_{i=0}^{L}p_i\lambda^i}  
{1+\sum_{j=1}^M q_j\lambda^j}=f(\lambda).
\end{eqnarray}
In our case the function $f(\lambda)$ represents the series for
$\frac{d}{d\lambda} 
\ln P_{\infty}(\lambda)$. As $f(\lambda)$ is a function of one variable, 
we fix the value of $\tilde{D}$ to calculate these approximants. For a fixed
value of $\tilde{D}$ one pole of the approximant $F$ will correspond to the
critical point while the associated residue will be the critical
exponent $\beta$.    
We calculate approximants with $L+M\leq 24$, restricting our
calculation to diagonal or close to diagonal approximants, which
usually display a better convergence. Thus
$L=M+\xi$, where $\xi=0,\pm 1$ and with $D=\tilde{D}/(1+\tilde{D})$
ranging between 0 and 0.8. Examples of
estimates for the critical values of $\mu$ obtained from these
approximants is given in Table \ref{tab11} for different values of 
the diffusion.

\begin{table}
\begin{center}
\begin{tabular}{cccc}
\hline
D&L&M&$\mu_c$\\
\hline
0  &10&10&0.60645\\
   &11&11&0.60646\\
0.1&10&10&0.62267\\
   &11&11&0.62266\\
0.7&10&10&0.85353\\
   &11&11&0.84513\\
0.8&10&10&0.96256\\
   &11&11&0.94246\\
\hline
\end{tabular}
\caption{Esimates for ritical points obtained by d-log Pad\'e
  approximants. Note that  
  as the value of $D=\tilde{D}/(1+\tilde{D})$ grows the dispersion in the
  value estimates also grows.}
\label{tab11}
\end{center}
\end{table}

\begin{figure}[h!]
\begin{center}
\vspace*{0.9cm}
\includegraphics[scale=0.5]{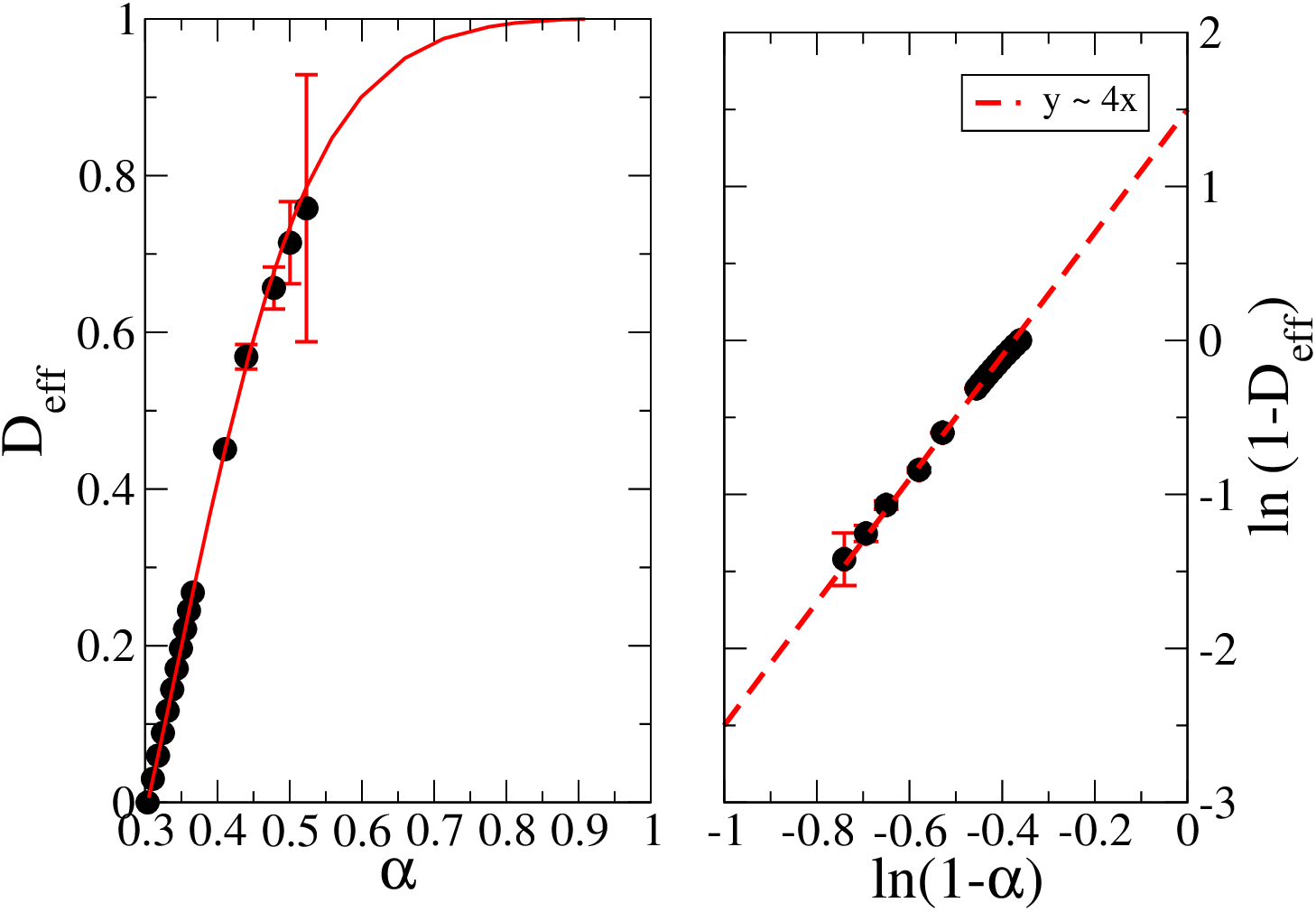}
\caption{At left panel we have the phase diagram obtained
by two-site mean-site approach. At right the log-log plot of the
same quantity is plotted showing that value of the crossover exponent $\phi$
is unity in this approach.}
\label{fig2}
\end{center}
\end{figure}

For each value of the diffusion rate, we calculate about eight 
approximants, obtaining the estimate of $\mu_c$ associated to
diffusion as an arithmetic average of results funnished by these set
of approximants.  
From this we obtain the phase diagram shown
in the figure \ref{fig2}. With the purpose of comparison 
with the results coming from the conservative simulations
\cite{mariofiore} 
we use the variables $\alpha\equiv\mu/2$ 
and $D_{eff}=\alpha\tilde{D}/(1+\alpha\tilde{D})$.  

For higher values of the diffusion the dispersion increases, 
and estimates with larger error bars are found.  Nevertheless, the
exponent $\phi=4$ seems to describe well the calculated points of the
critical line.  However, this result
would be different if we used approximants to series closer to  
the infinite diffusion rate limit.  
This error in the approximants for high values of the 
diffusion rate is attributed to the
alternated sign of the series terms \cite{dickjensdif}.  Another 
explanation would come from the fact that in the
neighborhood of a multicritical point the reduction of a 
two-variable series to one variable leads to very poor
estimates of the critical properties \cite{wgdstilckcross} close to a
multicritical point.
To overcome this problem we analyze the series 
using Partial Differential Approximants (PDA's) \cite{fk77}, 
that generalize the d-log Pad\'e approximants for a two-variable 
series.  These approximants are defined by the following equation  
\begin{equation}
P_{\mathbf L}(x,y)F(x,y)=Q_{\mathbf M}(x,y)\frac{\partial F(x,y)}{\partial x} +
R_{\mathbf N}(x,y) \frac{\partial F(x,y)}{\partial y},
\label{pda}
\end{equation}
where $P$, $Q$, and $R$ are polynomials in the variables $x$ and $y$ with the
set of nonzero coefficients ${\mathbf L}$, ${\mathbf M}$, and ${\mathbf N}$,
respectively. The coefficients 
of the polynomials are obtained by substitution of the series expansion
of the quantity which is going to be analyzed 
\begin{equation}
f(x,y)=\sum_{k,k^\prime=0} f(k,k^\prime)x^k y^{k^\prime}
\end{equation}
into equation (\ref{pda}) and requiring the equality to hold for a
set of indexes defined as ${\mathbf K}$. This procedure leads to a system
of linear 
equations for the coefficients of the polynomials, and since the coefficients
$f_{k,k^\prime}$ of the series are known for a finite set of indexes
this places 
an upper limit to the number of coefficients in the polynomials. Since the
number of equations has to match the number of unknown coefficients,
the numbers of elements in each set must satisfy $K=L+M+N-1$ (one
coefficient is fixed arbitrarily). An additional issue, 
which is not present in the one-variable case, is the symmetry of the
polynomials. Two frequently used options are the triangular and the
rectangular arrays of coefficients. The choice of these symmetries may
be related to the symmetry of the series itself \cite{s90}. 

It is possible to show \cite{s90} that we can  
determine the multicritical properties using the equation \ref{pda} 
and the hypothesis of that in the neighborhood of the multicritical point, 
the function $f$ obeys  the following scaling form
\begin{equation}
f(x,y) \approx |\Delta \widetilde{x}|^{-\nu}Z\left( \frac{|\Delta
\widetilde{y}|}{|\Delta \widetilde{x}|^\phi} \right),
\label{mcs}
\end{equation}
where
\begin{equation}
\Delta \widetilde{x}=(x-x_c )-(y-y_c)/e_2,
\end{equation}
and
\begin{equation}
\Delta \widetilde{y}=(y-y_c)-e_1(x-x_c).
\end{equation}
Here $\nu$ is the critical exponent of the quantity described by $f$ when
$\Delta \widetilde{y}=0$, $e_1$ and $e_2$ are the scaling slopes \cite{fk77}
and 
$\phi$ is the crossover exponent. 

\begin{figure}[h!]
\begin{center}
\vspace*{0.9cm}
\includegraphics[scale=0.5]{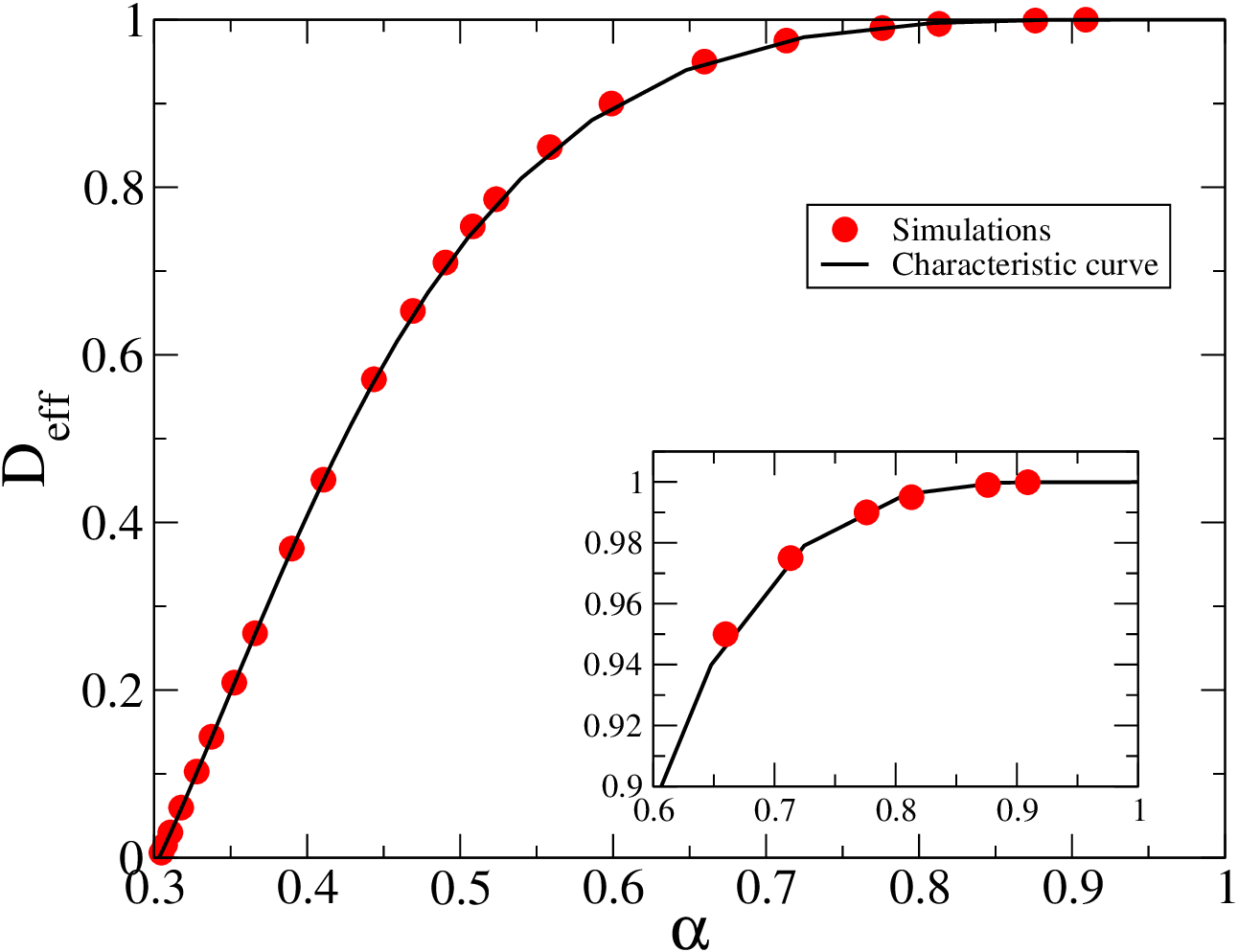}
\vspace*{0.2cm}
\caption{Comparision between a characteristic curve (solid line) and
a numerical simulation result (circles). The coincidence is evident,
including the region of the infinite diffusion limit (inset).}
\label{fig3}
\end{center}
\end{figure}

On the other side, our calculation was successful when we use 
the method of the characteristics to integrate equation 
(\ref{pda}).  This is made by introducing a timelike variable $\tau$, 
so that a family of curves is obtained in the plane 
$(x(\tau),y(\tau))$ (the characteristics).  
Such curves obey to the equations
\begin{eqnarray}
\frac{dx}{d \tau} &=& Q_M(x(\tau),y(\tau)),\nonumber\\
\frac{dy}{d\tau}  &=& R_N(x(\tau),y(\tau)).
\label{char}
\end{eqnarray}
It is possible to show that integrating the equations (\ref{char}) from a 
specific point of the critical line, the resulting 
characteristic is equivalent to the the critical line. In figure
\ref{fig3} we show a comparision between a characteristic curve and the 
simulational result \cite{mariofiore}. 
The number of elements in the sets of the calculated approximants was varying
as follows: $K=55-190$, $M=N=20-53$ and $L=15-54$.

In each of these curves, 
we calculate his inclination in the neighborhood of the multicritical
point, determining a value for the exponent $\phi$ and the mean value
for this exponent results as $\phi=4.02\pm0.13$, consistent with 
simulations in the particle conserving ensemble
\cite{mariofiore}. 

\begin{figure}[h!]
\begin{center}
\vspace*{0.9cm}
\includegraphics[scale=0.4]{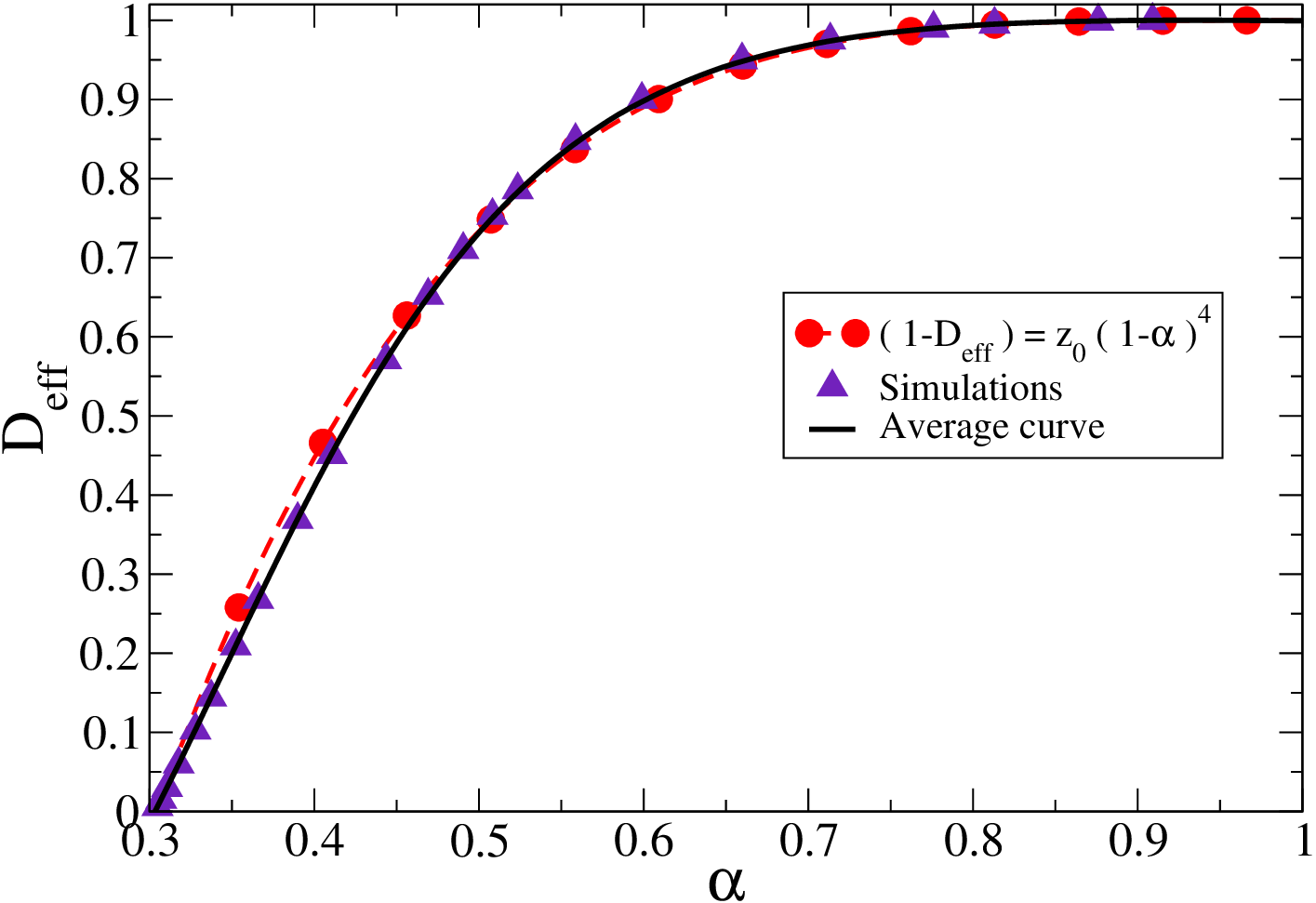}
\\
\vspace*{1.1cm}
\includegraphics[scale=0.4]{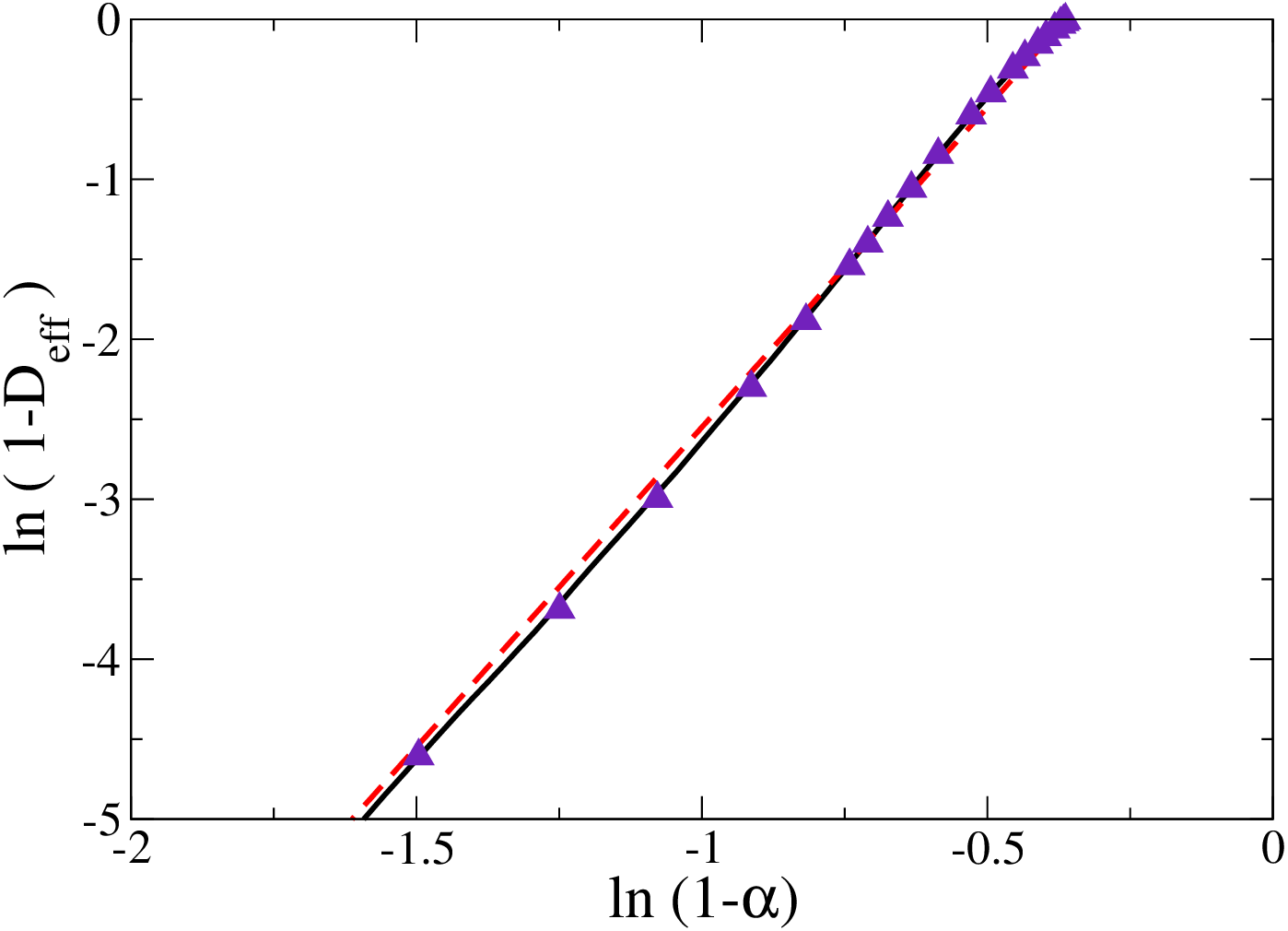}
\vspace*{0.2cm}
\caption{Comparision between a characteristic `average curve' (solid line), 
the numerical simulation result (triangles) and the curve obtained
from the scaling form $(1-D_{eff})\sim(1-\alpha)^4$(circles). 
At the right panel we see that $phi=4$ looks a good estimative for
the crossover exponent.}
\label{fig4}
\end{center}
\end{figure}

Using all the characteristic curves calculated we derive an `average curve', 
calculating for each value of $\alpha$ on arithmetic average for $D_{eff}$. 
This curve is shown in the figure \ref{fig4} jointly with the result
originating  
from the simulation and with the scaling form
$(1-D_{eff})\sim(1-\alpha)^4$.  In the same 
figure we see that the exponent $\phi=4$ is well fitted to the results of 
simulation and of the PDA's. This scaling form is based on the argument of the 
scaling function $Z$ presented in the equation (\ref{mcs}), where $\phi=4$ and
$z_0$ is a parameter properly chosen. We remark that this scaling form 
coincides with the characteristic curve and with the simulation even in the
weak diffusion regime. This is somewhat surprising since its validity
would be expected only in the neighborhood of the multicritical point.

Unfortunately, even using the algorithm proposed by Styer \cite{s90} 
we were not able to obtain precise estimates for 
the crossover exponent $\phi$ from the scaling form shown in 
equation (\ref{mcs}). However, integrating a set of
approximants, we could determine the characteristic curves   
whose initial point is conicident with the critical point of the CP 
without diffusion of particles. These curves are estimates 
for the critical line of the CP model with diffusion going 
beyond the values achieved in \cite{dickjensdif} 
and \cite{mariofiore} and corroborating the initial result of this last 
reference in that $\phi\approx 4$.
\section{Conclusion}
Calculating a supercritical series for the ultimate survival probability and 
analysing it using PDA's we obtain estimates for the critical line 
of the CP model with diffusion.  The critical line was derived through 
the integration of the equation (\ref{pda}) by the method 
of the characteristics.  Direct results for the value of the crossover 
exponent using the scaling form \ref{mcs} using Styer's algorithm  
\cite{s90} were not possible to be obtained with an acceptable precision.  
However the method of the characteristics permitted us to calculate
the critical 
line and, consequently, the value for the crossover exponent $\phi$.  
Our result, $\phi=4.02\pm0.13$ is in agreement with that derived 
in \cite{mariofiore} and explores a region of diffusion very close 
to the multicritical point.

The technique of the two-variable supercritical series associated 
with PDA analysis was shown to be accurate enough to determine the 
critical properties in similar models \cite{wgdstilckcross,mabdif}.  
Therefore, we believe that a natural extension for this work 
is analyze related models that apparently possess non-trivial 
multicritical points. This semms to be the cases of the pair-creation
and triplet-creation  
models with diffusion, also studied in \cite{mariofiore}.  
This research is already in course.  

W.G. Dantas acknowledges the financial 
support from  Funda\c{c}\~ao de Amparo \`a Pesquisa do
Estado de S\~ao Paulo (FAPESP) under Grant No. 05/04459-1.

\end{document}